\documentclass[showpacs,amsmath,superscriptaddress,
reprint,
aps,prb]{revtex4-1}

\usepackage{graphicx}
\begin{document}

\title
{Gate-induced blueshift and quenching of photoluminescence in suspended single-walled carbon nanotubes}
\author{S. Yasukochi}
\author{T. Murai}
\author{S. Moritsubo}
\author{T. Shimada}
\affiliation{Institute of Engineering Innovation, 
The University of Tokyo, Tokyo 113-8656, Japan}
\author{S. Chiashi}
\author{S. Maruyama}
\affiliation{Department of Mechanical Engineering, 
The University of Tokyo, Tokyo 113-8656, Japan}
\author{Y. K. Kato}
\email[Corresponding author. ]{ykato@sogo.t.u-tokyo.ac.jp}
\affiliation{Institute of Engineering Innovation, 
The University of Tokyo, Tokyo 113-8656, Japan}

\begin{abstract}
Gate-voltage effects on photoluminescence spectra of suspended single-walled carbon nanotubes are investigated. Photoluminescence microscopy and excitation spectroscopy are used to identify individual nanotubes and to determine their chiralities. Under an application of gate voltage, we observe slight blueshifts in the emission energy and strong quenching of photoluminescence. The blueshifts are similar for different chiralities investigated, suggesting extrinsic mechanisms. In addition, we find that the photoluminescence intensity quenches exponentially with gate voltage. 
\end{abstract}

\maketitle

Understanding of electric-field effects on optical emission from single-walled carbon nanotubes (SWCNTs) is a key to the development of carbon-based nanoscale optoelectronics.\cite{Avouris:2008} It has been shown that electric fields can drive light emission in SWCNTs.\cite{Misewich:2003, Mann:2007} In comparison, photoluminescence (PL) is quenched by an application of electric fields. Micelle-encapsulated SWCNTs show a reduction of PL intensity by electric fields along the tube axis.\cite{Naumov:2008} Similar quenching occurs in suspended nanotubes within field-effect transistor (FET) structures.\cite{Ohno:2006nanotech, Steiner:2009, Freitag:2009}  Such strong quenching of PL has made it difficult to unambiguously resolve shifts in the emission spectra, in contrast to  absorption measurements where redshifts due to the Stark effect \cite{Mohite:2008} and doping-induced screening \cite{Steiner:2009} have been observed.  
Detailed PL spectroscopy on nanotubes with determined chirality would help clarify the role of these effects on optical emission.

Here we report on gate-voltage dependence of PL spectra in individual suspended SWCNTs. As-grown  nanotubes within FET structures are identified by PL imaging using a laser scanning confocal microscope. Excitation spectroscopy is used to determine their chirality, and PL spectra are collected as a function of gate voltage. Surprisingly, we find that the emission blueshifts when the gate voltages are applied. Furthermore, the PL intensity decreases exponentially with gate voltage, and we find that a model assuming doping-induced exciton relaxation proportional to carrier density \cite{Perebeinos:2008,Kinder:2008} cannot account for all of the quenching observed.

\begin{figure}[b]
\includegraphics{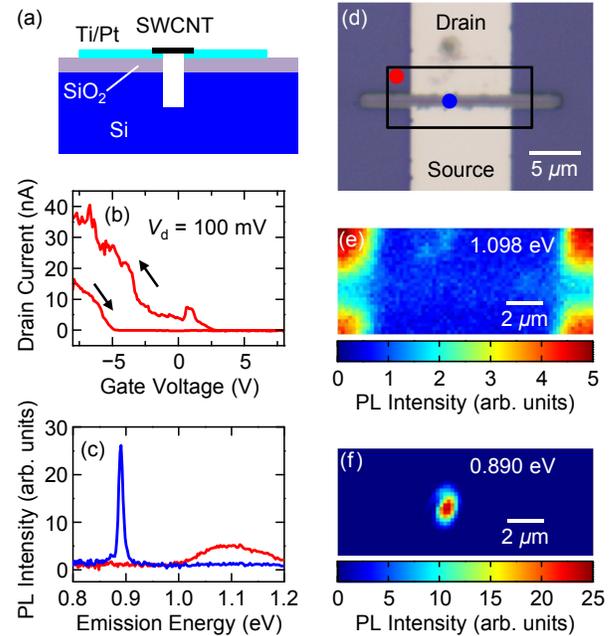}
\caption{\label{fig1}
(a) A schematic of a device.
(b) Typical electrical characteristics. Drain voltage $V_\text{d}=100$~mV is used.
(c) PL spectra for a (9,8)
carbon nanotube (blue) and Si substrate (red).
(d) Optical microscope image of a device.
The scale bar is 5~$\mu$m.
The blue and red dots indicate the positions of the laser spot
where the blue and red curves in (b) are taken, respectively.
The black box shows the scan area for the PL images.
(e) and (f) are PL images at emission energies
of 0.890~eV and 1.098~eV, respectively.
Spectral integration windows are 4~meV wide, and the scale bars are 2~$\mu$m.
For (c), (e), and (f), excitation energy of 1.653~eV and excitation power
of 0.65~mW are used.
}\end{figure}

The suspended nanotube FETs [Fig.~\ref{fig1}(a)] are fabricated on $p$-type Si substrates with 100-nm-thick oxide. We begin by etching 1-$\mu$m wide trenches with a depth of $\sim 5$~$ \mu$m. The wafer is then annealed at $900^\circ$C in oxygen for an hour to form an oxide layer inside the trenches.  1-nm Ti and 15-nm Pt are deposited for source and drain contacts, and then catalyst areas are defined on the drain electrodes. Co acetate and fumed silica are dissolved in ethanol, and deposited on the wafer. Finally, carbon nanotubes are grown by chemical vapor deposition using ethanol as a carbon source.\cite{Maruyama:2002} Typical device characteristics are shown in Fig.~\ref{fig1}(b), and we note that these devices show hysteresis due to water adsorption.\cite{Kim:2003}

PL spectra are collected with a home-built laser scanning confocal microscope.\cite{Moritsubo:2010} An output of a continuous-wave Ti:sapphire laser is focused onto the sample with an objective lens, and a steering mirror allows scanning of the laser spot. PL is detected with a liquid-nitrogen-cooled InGaAs photodiode array attached to a single-grating spectrometer. All measurements are done in air at room temperature.

Typical PL spectra are shown in Fig.~\ref{fig1}(c). The sharp peak at 0.9~eV originates from a suspended SWCNT, while PL from the Si substrate shows a broad emission at 1.1~eV. By scanning the laser spot and collecting PL spectra in the area shown in Fig.~\ref{fig1}(d), PL images for the Si substrate and the nanotube are constructed [Fig.~\ref{fig1}(e) and (f)]. Using these images, we check that the PL comes from a fully-suspended nanotube in between source and drain electrodes. Next, the chirality of the nanotube is identified by PL excitation spectroscopy,\cite{Bachilo:2002} using tabulated data for suspended SWCNTs.\cite{Ohno:2006prb} We look for nanotubes with a sharp single peak in the PL excitation map to exclude bundled tubes. 

\begin{figure}
\includegraphics{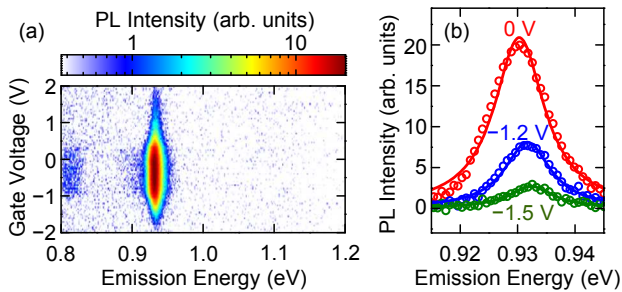}
\caption{\label{fig2}
(a) PL intensity as a function of gate voltage and emission energy for a (10,6) nanotube.
Excitation energy of 1.653~eV and excitation power of 126~$\mu$W are used. The source and drain contacts are grounded during these measurements.
(b) Typical spectra at selected gate voltages. Open circles are data and lines are Lorentzian fits. Red, blue, and green correspond to $V_\text{g}=0$~V, $-1.2$~V, and $-1.5$~V, respectively.
}\end{figure}

Following such procedures, we have investigated gate-field dependences of PL in 7 individual SWCNTs. Figure~\ref{fig2}(a) shows a series of PL spectra taken as a function of the gate voltage $V_\text{g}$, for a (10,6) nanotube.  The PL intensity is largest near $V_\text{g} = 0$ V, and decreases dramatically with an application of gate voltage, as observed previously.\cite{Ohno:2006nanotech, Steiner:2009} The PL intensity maximum is not exactly at $V_\text{g} = 0$ V, because of the hysteretic behavior of the devices. PL spectra at representative gate voltages are shown in Fig.~\ref{fig2}(b).  In addition to the quenching, we find that the emission blueshifts slightly as the gate voltage is applied. In order to quantitatively characterize the blueshift, we fit the PL peak with a Lorentzian function. The peak position and the peak area are extracted from the fitting parameters, and they are used as a measure of the emission energy and PL intensity, respectively.

\begin{figure*}
\includegraphics{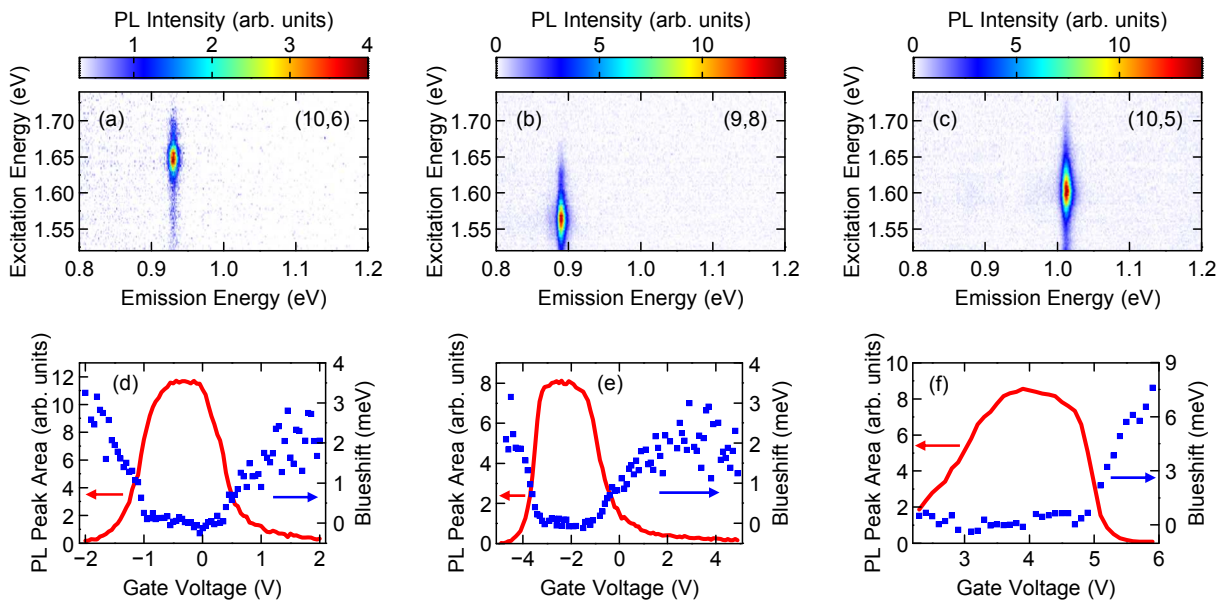}
\caption{\label{fig3}
(a--c) PL excitation maps for (10,6), (9,8), and (10,5) nanotubes, respectively.  The spectra are not corrected for the changes in excitation power as it does not vary by more than a factor of two within the tuning range. The nanotubes shown in (a) and (b) are the same as those shown in Fig.~\ref{fig2} and Fig.~\ref{fig1}(c--f), respectively. 
(d--f) Gate-voltage dependences of PL peak area (red line) and blueshift (blue squares) for the nanotubes shown in (a--c), respectively. In these measurements, the nanotubes are resonantly excited. Excitation powers of 126, 79, and 212~$\mu$W, and $V_\text{g}$ sweep rates of 18, 20, and 6~mV~s$^{-1}$ are used for (d--f), respectively. The gate voltage is scanned from negative to positive in panels (d) and (e), while it is scanned from positive to negative in panel (f). The blueshifts are measured from 930.4, 889.2, and 1008.2~meV in (d--f), respectively. 
}\end{figure*}

Such analysis is performed on all nanotubes investigated, and in Fig.~\ref{fig3} we present data from three nanotubes of different chirality.  For all three nanotubes, we observe the blueshift as well as quenching. The blueshift can be as much as 7~meV, and is limited by the loss of signal quality due to the quenching at higher gate voltages. 

We first consider the cause of the blueshift. Gate voltage may exert electrostatic forces on the nanotube and induce strain, resulting in changes in the bandgap energy that reverses its sign depending on the type of the nanotube.\cite{Yang:2000, Huang:2008} 
If strain is responsible for the observed energy shifts, we expect the sign of the energy shift for a (10,6) nanotube [Fig.~\ref{fig3}(a)] to be different from that of a (10,5) nanotube [Fig.~\ref{fig3}(c)]. Since we observe blueshifts regardless of the nanotube type, strain-induced changes in the bandgap cannot explain our data.

The relative insensitiveness of the blueshifts on chirality suggests extrinsic mechanisms. We speculate that the migration of molecules adsorbed on the surface of carbon nanotubes may be responsible for the blueshift. It has been shown that there exists a considerable amount of adsorbates on SWCNTs in air, and their removal causes a blueshift as much as 29~meV.\cite{Finnie:2005} In our FETs, the gate fields are expected to be concentrated near the source and drain contacts, and adsorbed molecules may migrate and get trapped in the high-field regions. The large hysteresis in gate sweeps is consistent with this interpretation, as it indicates charge trapping by the molecules.\cite{Kim:2003} Since we typically observe a blueshift of a few meV, reduction of the adsorbates by $\sim 10$\% is enough to explain our data. Further investigation in controlled environment would clarify the origin of the gate-induced blueshift.

The observation of blueshift is quite surprising as the Stark effect and screening by gate-induced carriers should cause redshifts.\cite{Mohite:2008, Steiner:2009} The absence of such redshifts implies that these effects do not play a role at these gate voltages for $E_{11}$ emission. This is reasonable for the Stark effect which requires significantly higher fields.\cite{Mohite:2008} Regarding the screening effect, redshifts for $E_{33}$ absorption at somewhat larger gate voltages has been observed using excitation energy dependence of Raman scattering.\cite{Steiner:2009} It is possible that screening becomes effective at higher gate voltages where PL is completely quenched, or that different transition bands behave differently.

We now turn our attention to the quenching of PL with gate voltage. Interpretation by carrier extraction \cite{Ohno:2006nanotech} has difficulties explaining exciton dissociation and carrier drift from the center of the trench to the contacts. Since fields perpendicular to the nanotube axis do not cause much quenching,\cite{Naumov:2008} it is likely that electrostatic doping plays an important role. Phase-space filling and doping-induced exciton relaxation have been suggested as possible mechanisms.\cite{Steiner:2009}

\begin{figure}
\includegraphics{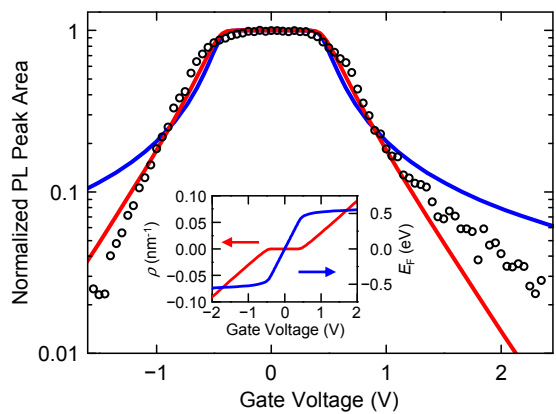}
\caption{\label{fig4}
 Normalized PL peak area as a function of $V_\text{g}$. The data (circles) are the same as in Fig.~\ref{fig3}(d), but an offset of 0.36~V have been added to $V_\text{g}$ to fit the simulation. Calculations assuming relaxation rates linear (blue curve) and exponential (red curve) in carrier density are plotted. $\alpha=1.30$~nm~ps$^{-1}$ is used for the blue curve, while $\beta=1.94\times 10^{-2}$~ps$^{-1}$ and  $\rho_0=2.48\times 10^{-2}$~nm$^{-1}$ are used for the red curve.
Inset shows calculated carrier density (red) and Fermi energy (blue) as a function of $V_\text{g}$. The hole densities are obtained by negating $V_\text{g}$, $E_\text{F}$, and $\rho$ in the calculations for electrons.
}\end{figure}

We model these doping-induced effects quantitatively and compare with the data for the (10,6) nanotube which has comparatively smaller hysteresis. We begin by computing the carrier density $\rho$ for a Fermi energy $E_\text{F}$ from $\rho (E_\text{F})= \int_{E_1}^{\infty} dE \left[g(E) f(E, E_\text{F})\right]$,
where $g(E)=\frac{4}{\pi \hbar v_\text{F}} \frac{E}{\sqrt{E^2-E_1^2}}$
is the density of states for the first conduction band at an electronic energy $E$, $\hbar$ is the Planck constant, $v_\text{F}= 9.5 \times 10^5$~m~s$^{-1}$ is the Fermi velocity in graphene, $E_1=(E_{11}+E_\text{b})/2$ is half the bandgap energy,  $E_{11}=0.930$~eV is the excitonic transition energy, $E_\text{b}=0.34/d$~eV is the exciton binding energy,\cite{Dukovic:2005} $d=1.1$~nm is the nanotube diameter,
$f(E, E_\text{F})=[1+\exp (\frac{E-E_\text{F}}{ k_\text{B} T})]^{-1}$
is the Fermi-Dirac distribution function, $k_\text{B}$ is the Boltzmann constant, and $T=300$~K is the temperature. Taking into account the quantum capacitance,\cite{Ilani:2006} we obtain the relation between $V_\text{g}$ and $E_\text{F}$ from $V_\text{g} (E_\text{F})=\frac{E_\text{F}}{e}+\frac{e\rho(E_\text{F})}{C_\text{g}}$,
where $e$ is the electronic charge and $C_\text{g}$ is the geometric capacitance of the gate. We use $C_\text{g}=10$~aF~$\mu$m$^{-1}$, based on estimates in similar devices.\cite{Bushmaker:2009, Steiner:2009}
The results of the calculation are shown in the inset of Fig.~\ref{fig4}.

To simulate the changes in the PL efficiency, we calculate $\gamma_0/[\gamma_0+\gamma_\text{d}(\rho)]$ where $\gamma_0 = 0.01$~ps$^{-1}$ is the exciton relaxation rate in the absence of doping \cite{Xiao:2010} and $\gamma_\text{d}(\rho)$ is the doping-induced relaxation rate. Assuming $\gamma_\text{d}$ proportional to $\rho$,\cite{Perebeinos:2008,Kinder:2008,Matsuda:2010} we have performed a least-squares fit to normalized data with $\gamma_\text{d}=\alpha \rho$ where $\alpha$ is a proportionality constant used as a fitting parameter (Fig.~\ref{fig4}, blue curve).  

There are several issues with this doping-induced relaxation model. First of all, we obtain $\alpha=1.30$~nm~ps$^{-1}$ which is about an order of magnitude larger than the theoretical calculation.\cite{Perebeinos:2008} Secondly, it does not reproduce the quenching exponential with $V_\text{g}$. In addition, at higher gate voltages, the model deviates from the experimental data by factors of 2 to 5.

As a possible reason for this discrepancy, we consider the phase-space filling effect.\cite{Luer:2009,Steiner:2009}  For an exciton with zero center-of-mass momentum, the wave function $\Psi$ in $k$-space can be written as $\Psi(k)=\exp(-k^2\sigma^2/2)$ where $k$ is the relative momentum and $\sigma$ is the exciton radius.\cite{Luer:2009} We assume that electron and hole masses are the same, so that the relative momentum of exciton is equal to the electron momentum. Using the electron dispersion $E^2=E_1^2+\hbar^2 v_\text{F}^2 k^2$, the fraction of the filled states is given by
\[
\frac{
\int_{E_1}^{\infty} dE \left[\exp(-\frac{E^2-E_1^2}{2 \hbar^2 v_\text{F}^2}\sigma^2) g(E) f(E, E_\text{F})\right]
}{
\int_{E_1}^{\infty} dE \left[\exp(-\frac{E^2-E_1^2}{2 \hbar^2 v_\text{F}^2}\sigma^2) g(E)\right]
},
\]
but it can only account for $\sim10$\% reduction in the PL intensity at $V_\text{g}=2$~V for $\sigma=2$~nm. As another possibility, $E_{22}$ may also blueshift along with $E_{11}$ and change the absorption efficiency, but $E_{22}$ linewidth of $\sim 40$~meV is too broad to account for the descrepancy.

To better describe the observed results, we have introduced a phenomenological expression for doping-induced relaxation given by $\gamma_\text{d}=\beta [\exp(|\rho|/\rho_0)-1]$ and performed a least-squares fit with $\beta$ and $\rho_0$ as fitting parameters (red curve in Fig.~\ref{fig4}). Such an expression implies exponential increase of relaxation rate with carrier density. Since $\rho$ increases almost linearly with $V_\text{g}$, mechanisms that scale exponentially with gate voltages or fields are also plausible. Yet another possibility is the role of carriers in higher bands, which would scale exponentially with $V_\text{g}$ because of the tail of the Fermi-Dirac distribution.

In addition to the bright exciton emission that we have discussed, there also exists a weak emission at a lower energy.  In Fig.~\ref{fig2}(a), there is a small peak around 0.8~eV due to $K$-momentum excitons.\cite{Torrens:2008, Murakami:2009, Matsunaga:2010} This peak also disappears upon application of the gate voltage, consistent with the prediction that $K$-momentum excitons also experience doping-induced relaxation.\cite{Perebeinos:2008}

In summary, we have investigated the gate-field effect on PL spectra of individual suspended SWCNTs with determined chiralities. Slight blueshifts due to applied gate-fields have been observed for three different chiralities, suggesting extrinsic mechanisms. The lack of any redshift in PL implies that screening effects are ineffective within the gate fields investigated. Quenching increases exponentially with gate voltage, and a comparison with theoretical model shows that doping-induced exciton relaxation may be stronger than previously thought.

\begin{acknowledgments}
We thank T. Kan and I. Shimoyama for the use of electron beam evaporator, and acknowledge support from KAKENHI, Mizuho Foundation for the Promotion of Sciences, Research Foundation for Opto-Science and Technology, SCAT, SCOPE, TEPCO Memorial Foundation, The Japan Prize Foundation, and Photon Frontier Network Program of MEXT, Japan. The devices were fabricated at the Center for Nano Lithography \& Analysis at The University of Tokyo.
\end{acknowledgments}

\bibliography{Blueshift}

\begin{thebibliography}{27}%
\makeatletter
\providecommand \@ifxundefined [1]{%
 \@ifx{#1\undefined}
}%
\providecommand \@ifnum [1]{%
 \ifnum #1\expandafter \@firstoftwo
 \else \expandafter \@secondoftwo
 \fi
}%
\providecommand \@ifx [1]{%
 \ifx #1\expandafter \@firstoftwo
 \else \expandafter \@secondoftwo
 \fi
}%
\providecommand \natexlab [1]{#1}%
\providecommand \enquote  [1]{``#1''}%
\providecommand \bibnamefont  [1]{#1}%
\providecommand \bibfnamefont [1]{#1}%
\providecommand \citenamefont [1]{#1}%
\providecommand \href@noop [0]{\@secondoftwo}%
\providecommand \href [0]{\begingroup \@sanitize@url \@href}%
\providecommand \@href[1]{\@@startlink{#1}\@@href}%
\providecommand \@@href[1]{\endgroup#1\@@endlink}%
\providecommand \@sanitize@url [0]{\catcode `\\12\catcode `\$12\catcode
  `\&12\catcode `\#12\catcode `\^12\catcode `\_12\catcode `\%12\relax}%
\providecommand \@@startlink[1]{}%
\providecommand \@@endlink[0]{}%
\providecommand \url  [0]{\begingroup\@sanitize@url \@url }%
\providecommand \@url [1]{\endgroup\@href {#1}{\urlprefix }}%
\providecommand \urlprefix  [0]{URL }%
\providecommand \Eprint [0]{\href }%
\providecommand \doibase [0]{http://dx.doi.org/}%
\providecommand \selectlanguage [0]{\@gobble}%
\providecommand \bibinfo  [0]{\@secondoftwo}%
\providecommand \bibfield  [0]{\@secondoftwo}%
\providecommand \translation [1]{[#1]}%
\providecommand \BibitemOpen [0]{}%
\providecommand \bibitemStop [0]{}%
\providecommand \bibitemNoStop [0]{.\EOS\space}%
\providecommand \EOS [0]{\spacefactor3000\relax}%
\providecommand \BibitemShut  [1]{\csname bibitem#1\endcsname}%
\let\auto@bib@innerbib\@empty
\bibitem [{\citenamefont {Avouris}\ \emph {et~al.}(2008)\citenamefont
  {Avouris}, \citenamefont {Freitag},\ and\ \citenamefont
  {Perebeinos}}]{Avouris:2008}%
  \BibitemOpen
  \bibfield  {author} {\bibinfo {author} {\bibfnamefont {P.}~\bibnamefont
  {Avouris}}, \bibinfo {author} {\bibfnamefont {M.}~\bibnamefont {Freitag}}, \
  and\ \bibinfo {author} {\bibfnamefont {V.}~\bibnamefont {Perebeinos}},\
  }\href {\doibase 10.1038/nphoton.2008.94} {\bibfield  {journal} {\bibinfo
  {journal} {Nature Photon.}\ }\textbf {\bibinfo {volume} {2}},\ \bibinfo
  {pages} {341} (\bibinfo {year} {2008})}\BibitemShut {NoStop}%
\bibitem [{\citenamefont {Misewich}\ \emph {et~al.}(2003)\citenamefont
  {Misewich}, \citenamefont {Martel}, \citenamefont {Avouris}, \citenamefont
  {Tsang}, \citenamefont {Heinze},\ and\ \citenamefont
  {Tersoff}}]{Misewich:2003}%
  \BibitemOpen
  \bibfield  {author} {\bibinfo {author} {\bibfnamefont {J.~A.}\ \bibnamefont
  {Misewich}}, \bibinfo {author} {\bibfnamefont {R.}~\bibnamefont {Martel}},
  \bibinfo {author} {\bibfnamefont {P.}~\bibnamefont {Avouris}}, \bibinfo
  {author} {\bibfnamefont {J.~C.}\ \bibnamefont {Tsang}}, \bibinfo {author}
  {\bibfnamefont {S.}~\bibnamefont {Heinze}}, \ and\ \bibinfo {author}
  {\bibfnamefont {J.}~\bibnamefont {Tersoff}},\ }\href {\doibase
  10.1126/science.1081294} {\bibfield  {journal} {\bibinfo  {journal}
  {Science}\ }\textbf {\bibinfo {volume} {300}},\ \bibinfo {pages} {783}
  (\bibinfo {year} {2003})}\BibitemShut {NoStop}%
\bibitem [{\citenamefont {Mann}\ \emph {et~al.}(2007)\citenamefont {Mann},
  \citenamefont {Kato}, \citenamefont {Kinkhabwala}, \citenamefont {Pop},
  \citenamefont {Cao}, \citenamefont {Wang}, \citenamefont {Zhang},
  \citenamefont {Wang}, \citenamefont {Guo},\ and\ \citenamefont
  {Dai}}]{Mann:2007}%
  \BibitemOpen
  \bibfield  {author} {\bibinfo {author} {\bibfnamefont {D.}~\bibnamefont
  {Mann}}, \bibinfo {author} {\bibfnamefont {Y.~K.}\ \bibnamefont {Kato}},
  \bibinfo {author} {\bibfnamefont {A.}~\bibnamefont {Kinkhabwala}}, \bibinfo
  {author} {\bibfnamefont {E.}~\bibnamefont {Pop}}, \bibinfo {author}
  {\bibfnamefont {J.}~\bibnamefont {Cao}}, \bibinfo {author} {\bibfnamefont
  {X.}~\bibnamefont {Wang}}, \bibinfo {author} {\bibfnamefont {L.}~\bibnamefont
  {Zhang}}, \bibinfo {author} {\bibfnamefont {Q.}~\bibnamefont {Wang}},
  \bibinfo {author} {\bibfnamefont {J.}~\bibnamefont {Guo}}, \ and\ \bibinfo
  {author} {\bibfnamefont {H.}~\bibnamefont {Dai}},\ }\href {\doibase
  10.1038/nnano.2006.169} {\bibfield  {journal} {\bibinfo  {journal} {Nature
  Nanotech.}\ }\textbf {\bibinfo {volume} {2}},\ \bibinfo {pages} {33}
  (\bibinfo {year} {2007})}\BibitemShut {NoStop}%
\bibitem [{\citenamefont {Naumov}\ \emph {et~al.}(2008)\citenamefont {Naumov},
  \citenamefont {Bachilo}, \citenamefont {Tsyboulski},\ and\ \citenamefont
  {Weisman}}]{Naumov:2008}%
  \BibitemOpen
  \bibfield  {author} {\bibinfo {author} {\bibfnamefont {A.~V.}\ \bibnamefont
  {Naumov}}, \bibinfo {author} {\bibfnamefont {S.~M.}\ \bibnamefont {Bachilo}},
  \bibinfo {author} {\bibfnamefont {D.~A.}\ \bibnamefont {Tsyboulski}}, \ and\
  \bibinfo {author} {\bibfnamefont {R.~B.}\ \bibnamefont {Weisman}},\ }\href
  {\doibase 10.1021/nl0800974} {\bibfield  {journal} {\bibinfo  {journal} {Nano
  Lett.}\ }\textbf {\bibinfo {volume} {8}},\ \bibinfo {pages} {1527} (\bibinfo
  {year} {2008})}\BibitemShut {NoStop}%
\bibitem [{\citenamefont {Ohno}\ \emph
  {et~al.}(2006{\natexlab{a}})\citenamefont {Ohno}, \citenamefont {Kishimoto},\
  and\ \citenamefont {Mizutani}}]{Ohno:2006nanotech}%
  \BibitemOpen
  \bibfield  {author} {\bibinfo {author} {\bibfnamefont {Y.}~\bibnamefont
  {Ohno}}, \bibinfo {author} {\bibfnamefont {S.}~\bibnamefont {Kishimoto}}, \
  and\ \bibinfo {author} {\bibfnamefont {T.}~\bibnamefont {Mizutani}},\ }\href
  {\doibase 10.1088/0957-4484/17/2/035} {\bibfield  {journal} {\bibinfo
  {journal} {Nanotechnology}\ }\textbf {\bibinfo {volume} {17}},\ \bibinfo
  {pages} {549} (\bibinfo {year} {2006}{\natexlab{a}})}\BibitemShut {NoStop}%
\bibitem [{\citenamefont {Steiner}\ \emph {et~al.}(2009)\citenamefont
  {Steiner}, \citenamefont {Freitag}, \citenamefont {Perebeinos}, \citenamefont
  {Naumov}, \citenamefont {Small}, \citenamefont {Bol},\ and\ \citenamefont
  {Avouris}}]{Steiner:2009}%
  \BibitemOpen
  \bibfield  {author} {\bibinfo {author} {\bibfnamefont {M.}~\bibnamefont
  {Steiner}}, \bibinfo {author} {\bibfnamefont {M.}~\bibnamefont {Freitag}},
  \bibinfo {author} {\bibfnamefont {V.}~\bibnamefont {Perebeinos}}, \bibinfo
  {author} {\bibfnamefont {A.}~\bibnamefont {Naumov}}, \bibinfo {author}
  {\bibfnamefont {J.~P.}\ \bibnamefont {Small}}, \bibinfo {author}
  {\bibfnamefont {A.~A.}\ \bibnamefont {Bol}}, \ and\ \bibinfo {author}
  {\bibfnamefont {P.}~\bibnamefont {Avouris}},\ }\href {\doibase
  10.1021/nl9016804} {\bibfield  {journal} {\bibinfo  {journal} {Nano Lett.}\
  }\textbf {\bibinfo {volume} {9}},\ \bibinfo {pages} {3477} (\bibinfo {year}
  {2009})}\BibitemShut {NoStop}%
\bibitem [{\citenamefont {Freitag}\ \emph {et~al.}(2009)\citenamefont
  {Freitag}, \citenamefont {Steiner}, \citenamefont {Naumov}, \citenamefont
  {Small}, \citenamefont {Bol}, \citenamefont {Perebeinos},\ and\ \citenamefont
  {Avouris}}]{Freitag:2009}%
  \BibitemOpen
  \bibfield  {author} {\bibinfo {author} {\bibfnamefont {M.}~\bibnamefont
  {Freitag}}, \bibinfo {author} {\bibfnamefont {M.}~\bibnamefont {Steiner}},
  \bibinfo {author} {\bibfnamefont {A.}~\bibnamefont {Naumov}}, \bibinfo
  {author} {\bibfnamefont {J.~P.}\ \bibnamefont {Small}}, \bibinfo {author}
  {\bibfnamefont {A.~A.}\ \bibnamefont {Bol}}, \bibinfo {author} {\bibfnamefont
  {V.}~\bibnamefont {Perebeinos}}, \ and\ \bibinfo {author} {\bibfnamefont
  {P.}~\bibnamefont {Avouris}},\ }\href {\doibase 10.1021/nn900962f} {\bibfield
   {journal} {\bibinfo  {journal} {ACS Nano}\ }\textbf {\bibinfo {volume}
  {3}},\ \bibinfo {pages} {3744} (\bibinfo {year} {2009})}\BibitemShut
  {NoStop}%
\bibitem [{\citenamefont {Mohite}\ \emph {et~al.}(2008)\citenamefont {Mohite},
  \citenamefont {Gopinath}, \citenamefont {Shah},\ and\ \citenamefont
  {Alphenaar}}]{Mohite:2008}%
  \BibitemOpen
  \bibfield  {author} {\bibinfo {author} {\bibfnamefont {A.~D.}\ \bibnamefont
  {Mohite}}, \bibinfo {author} {\bibfnamefont {P.}~\bibnamefont {Gopinath}},
  \bibinfo {author} {\bibfnamefont {H.~M.}\ \bibnamefont {Shah}}, \ and\
  \bibinfo {author} {\bibfnamefont {B.~W.}\ \bibnamefont {Alphenaar}},\ }\href
  {\doibase 10.1021/nl0722525} {\bibfield  {journal} {\bibinfo  {journal} {Nano
  Lett.}\ }\textbf {\bibinfo {volume} {8}},\ \bibinfo {pages} {142} (\bibinfo
  {year} {2008})}\BibitemShut {NoStop}%
\bibitem [{\citenamefont {Perebeinos}\ and\ \citenamefont
  {Avouris}(2008)}]{Perebeinos:2008}%
  \BibitemOpen
  \bibfield  {author} {\bibinfo {author} {\bibfnamefont {V.}~\bibnamefont
  {Perebeinos}}\ and\ \bibinfo {author} {\bibfnamefont {P.}~\bibnamefont
  {Avouris}},\ }\href {\doibase 10.1103/PhysRevLett.101.057401} {\bibfield
  {journal} {\bibinfo  {journal} {Phys. Rev. Lett.}\ }\textbf {\bibinfo
  {volume} {101}},\ \bibinfo {pages} {057401} (\bibinfo {year}
  {2008})}\BibitemShut {NoStop}%
\bibitem [{\citenamefont {Kinder}\ and\ \citenamefont
  {Mele}(2008)}]{Kinder:2008}%
  \BibitemOpen
  \bibfield  {author} {\bibinfo {author} {\bibfnamefont {J.~M.}\ \bibnamefont
  {Kinder}}\ and\ \bibinfo {author} {\bibfnamefont {E.~J.}\ \bibnamefont
  {Mele}},\ }\href {\doibase 10.1103/PhysRevB.78.155429} {\bibfield  {journal}
  {\bibinfo  {journal} {Phys. Rev. B}\ }\textbf {\bibinfo {volume} {78}},\
  \bibinfo {pages} {155429} (\bibinfo {year} {2008})}\BibitemShut {NoStop}%
\bibitem [{\citenamefont {Maruyama}\ \emph {et~al.}(2002)\citenamefont
  {Maruyama}, \citenamefont {Kojima}, \citenamefont {Miyauchi}, \citenamefont
  {Chiashi},\ and\ \citenamefont {Kohno}}]{Maruyama:2002}%
  \BibitemOpen
  \bibfield  {author} {\bibinfo {author} {\bibfnamefont {S.}~\bibnamefont
  {Maruyama}}, \bibinfo {author} {\bibfnamefont {R.}~\bibnamefont {Kojima}},
  \bibinfo {author} {\bibfnamefont {Y.}~\bibnamefont {Miyauchi}}, \bibinfo
  {author} {\bibfnamefont {S.}~\bibnamefont {Chiashi}}, \ and\ \bibinfo
  {author} {\bibfnamefont {M.}~\bibnamefont {Kohno}},\ }\href {\doibase
  10.1016/S0009-2614(02)00838-2} {\bibfield  {journal} {\bibinfo  {journal}
  {Chem. Phys. Lett.}\ }\textbf {\bibinfo {volume} {360}},\ \bibinfo {pages}
  {229} (\bibinfo {year} {2002})}\BibitemShut {NoStop}%
\bibitem [{\citenamefont {Kim}\ \emph {et~al.}(2003)\citenamefont {Kim},
  \citenamefont {Javey}, \citenamefont {Vermesh}, \citenamefont {Wang},
  \citenamefont {Li},\ and\ \citenamefont {Dai}}]{Kim:2003}%
  \BibitemOpen
  \bibfield  {author} {\bibinfo {author} {\bibfnamefont {W.}~\bibnamefont
  {Kim}}, \bibinfo {author} {\bibfnamefont {A.}~\bibnamefont {Javey}}, \bibinfo
  {author} {\bibfnamefont {O.}~\bibnamefont {Vermesh}}, \bibinfo {author}
  {\bibfnamefont {Q.}~\bibnamefont {Wang}}, \bibinfo {author} {\bibfnamefont
  {Y.}~\bibnamefont {Li}}, \ and\ \bibinfo {author} {\bibfnamefont
  {H.}~\bibnamefont {Dai}},\ }\href {\doibase 10.1021/nl0259232} {\bibfield
  {journal} {\bibinfo  {journal} {Nano Lett.}\ }\textbf {\bibinfo {volume}
  {3}},\ \bibinfo {pages} {193} (\bibinfo {year} {2003})}\BibitemShut {NoStop}%
\bibitem [{\citenamefont {Moritsubo}\ \emph {et~al.}(2010)\citenamefont
  {Moritsubo}, \citenamefont {Murai}, \citenamefont {Shimada}, \citenamefont
  {Murakami}, \citenamefont {Chiashi}, \citenamefont {Maruyama},\ and\
  \citenamefont {Kato}}]{Moritsubo:2010}%
  \BibitemOpen
  \bibfield  {author} {\bibinfo {author} {\bibfnamefont {S.}~\bibnamefont
  {Moritsubo}}, \bibinfo {author} {\bibfnamefont {T.}~\bibnamefont {Murai}},
  \bibinfo {author} {\bibfnamefont {T.}~\bibnamefont {Shimada}}, \bibinfo
  {author} {\bibfnamefont {Y.}~\bibnamefont {Murakami}}, \bibinfo {author}
  {\bibfnamefont {S.}~\bibnamefont {Chiashi}}, \bibinfo {author} {\bibfnamefont
  {S.}~\bibnamefont {Maruyama}}, \ and\ \bibinfo {author} {\bibfnamefont
  {Y.~K.}\ \bibnamefont {Kato}},\ }\href {\doibase
  10.1103/PhysRevLett.104.247402} {\bibfield  {journal} {\bibinfo  {journal}
  {Phys. Rev. Lett.}\ }\textbf {\bibinfo {volume} {104}},\ \bibinfo {pages}
  {247402} (\bibinfo {year} {2010})}\BibitemShut {NoStop}%
\bibitem [{\citenamefont {Bachilo}\ \emph {et~al.}(2002)\citenamefont
  {Bachilo}, \citenamefont {Strano}, \citenamefont {Kittrell}, \citenamefont
  {Hauge}, \citenamefont {Smalley},\ and\ \citenamefont
  {Weisman}}]{Bachilo:2002}%
  \BibitemOpen
  \bibfield  {author} {\bibinfo {author} {\bibfnamefont {S.~M.}\ \bibnamefont
  {Bachilo}}, \bibinfo {author} {\bibfnamefont {M.~S.}\ \bibnamefont {Strano}},
  \bibinfo {author} {\bibfnamefont {C.}~\bibnamefont {Kittrell}}, \bibinfo
  {author} {\bibfnamefont {R.~H.}\ \bibnamefont {Hauge}}, \bibinfo {author}
  {\bibfnamefont {R.~E.}\ \bibnamefont {Smalley}}, \ and\ \bibinfo {author}
  {\bibfnamefont {R.~B.}\ \bibnamefont {Weisman}},\ }\href {\doibase
  10.1126/science.1078727} {\bibfield  {journal} {\bibinfo  {journal}
  {Science}\ }\textbf {\bibinfo {volume} {298}},\ \bibinfo {pages} {2361}
  (\bibinfo {year} {2002})}\BibitemShut {NoStop}%
\bibitem [{\citenamefont {Ohno}\ \emph
  {et~al.}(2006{\natexlab{b}})\citenamefont {Ohno}, \citenamefont {Iwasaki},
  \citenamefont {Murakami}, \citenamefont {Kishimoto}, \citenamefont
  {Maruyama},\ and\ \citenamefont {Mizutani}}]{Ohno:2006prb}%
  \BibitemOpen
  \bibfield  {author} {\bibinfo {author} {\bibfnamefont {Y.}~\bibnamefont
  {Ohno}}, \bibinfo {author} {\bibfnamefont {S.}~\bibnamefont {Iwasaki}},
  \bibinfo {author} {\bibfnamefont {Y.}~\bibnamefont {Murakami}}, \bibinfo
  {author} {\bibfnamefont {S.}~\bibnamefont {Kishimoto}}, \bibinfo {author}
  {\bibfnamefont {S.}~\bibnamefont {Maruyama}}, \ and\ \bibinfo {author}
  {\bibfnamefont {T.}~\bibnamefont {Mizutani}},\ }\href {\doibase
  10.1103/PhysRevB.73.235427} {\bibfield  {journal} {\bibinfo  {journal} {Phys.
  Rev. B}\ }\textbf {\bibinfo {volume} {73}},\ \bibinfo {pages} {235427}
  (\bibinfo {year} {2006}{\natexlab{b}})}\BibitemShut {NoStop}%
\bibitem [{\citenamefont {Yang}\ and\ \citenamefont {Han}(2000)}]{Yang:2000}%
  \BibitemOpen
  \bibfield  {author} {\bibinfo {author} {\bibfnamefont {L.}~\bibnamefont
  {Yang}}\ and\ \bibinfo {author} {\bibfnamefont {J.}~\bibnamefont {Han}},\
  }\href {\doibase 10.1103/PhysRevLett.85.154} {\bibfield  {journal} {\bibinfo
  {journal} {Phys. Rev. Lett.}\ }\textbf {\bibinfo {volume} {85}},\ \bibinfo
  {pages} {154} (\bibinfo {year} {2000})}\BibitemShut {NoStop}%
\bibitem [{\citenamefont {Huang}\ \emph {et~al.}(2008)\citenamefont {Huang},
  \citenamefont {Wu}, \citenamefont {Chandra}, \citenamefont {Yan},
  \citenamefont {Shan}, \citenamefont {Heinz},\ and\ \citenamefont
  {Hone}}]{Huang:2008}%
  \BibitemOpen
  \bibfield  {author} {\bibinfo {author} {\bibfnamefont {M.}~\bibnamefont
  {Huang}}, \bibinfo {author} {\bibfnamefont {Y.}~\bibnamefont {Wu}}, \bibinfo
  {author} {\bibfnamefont {B.}~\bibnamefont {Chandra}}, \bibinfo {author}
  {\bibfnamefont {H.}~\bibnamefont {Yan}}, \bibinfo {author} {\bibfnamefont
  {Y.}~\bibnamefont {Shan}}, \bibinfo {author} {\bibfnamefont {T.~F.}\
  \bibnamefont {Heinz}}, \ and\ \bibinfo {author} {\bibfnamefont
  {J.}~\bibnamefont {Hone}},\ }\href {\doibase 10.1103/PhysRevLett.100.136803}
  {\bibfield  {journal} {\bibinfo  {journal} {Phys. Rev. Lett.}\ }\textbf
  {\bibinfo {volume} {100}},\ \bibinfo {pages} {136803} (\bibinfo {year}
  {2008})}\BibitemShut {NoStop}%
\bibitem [{\citenamefont {Finnie}\ \emph {et~al.}(2005)\citenamefont {Finnie},
  \citenamefont {Homma},\ and\ \citenamefont {Lefebvre}}]{Finnie:2005}%
  \BibitemOpen
  \bibfield  {author} {\bibinfo {author} {\bibfnamefont {P.}~\bibnamefont
  {Finnie}}, \bibinfo {author} {\bibfnamefont {Y.}~\bibnamefont {Homma}}, \
  and\ \bibinfo {author} {\bibfnamefont {J.}~\bibnamefont {Lefebvre}},\ }\href
  {\doibase 10.1103/PhysRevLett.94.247401} {\bibfield  {journal} {\bibinfo
  {journal} {Phys. Rev. Lett.}\ }\textbf {\bibinfo {volume} {94}},\ \bibinfo
  {pages} {247401} (\bibinfo {year} {2005})}\BibitemShut {NoStop}%
\bibitem [{\citenamefont {Dukovic}\ \emph {et~al.}(2005)\citenamefont
  {Dukovic}, \citenamefont {Wang}, \citenamefont {Song}, \citenamefont {Sfeir},
  \citenamefont {Heinz},\ and\ \citenamefont {Brus}}]{Dukovic:2005}%
  \BibitemOpen
  \bibfield  {author} {\bibinfo {author} {\bibfnamefont {G.}~\bibnamefont
  {Dukovic}}, \bibinfo {author} {\bibfnamefont {F.}~\bibnamefont {Wang}},
  \bibinfo {author} {\bibfnamefont {D.}~\bibnamefont {Song}}, \bibinfo {author}
  {\bibfnamefont {M.~Y.}\ \bibnamefont {Sfeir}}, \bibinfo {author}
  {\bibfnamefont {T.~F.}\ \bibnamefont {Heinz}}, \ and\ \bibinfo {author}
  {\bibfnamefont {L.~E.}\ \bibnamefont {Brus}},\ }\href {\doibase
  10.1021/nl0518122} {\bibfield  {journal} {\bibinfo  {journal} {Nano Lett.}\
  }\textbf {\bibinfo {volume} {5}},\ \bibinfo {pages} {2314} (\bibinfo {year}
  {2005})}\BibitemShut {NoStop}%
\bibitem [{\citenamefont {Ilani}\ \emph {et~al.}(2006)\citenamefont {Ilani},
  \citenamefont {Donev}, \citenamefont {Kindermann},\ and\ \citenamefont
  {McEuen}}]{Ilani:2006}%
  \BibitemOpen
  \bibfield  {author} {\bibinfo {author} {\bibfnamefont {S.}~\bibnamefont
  {Ilani}}, \bibinfo {author} {\bibfnamefont {L.~A.~K.}\ \bibnamefont {Donev}},
  \bibinfo {author} {\bibfnamefont {M.}~\bibnamefont {Kindermann}}, \ and\
  \bibinfo {author} {\bibfnamefont {P.~L.}\ \bibnamefont {McEuen}},\ }\href
  {\doibase 10.1038/nphys412} {\bibfield  {journal} {\bibinfo  {journal}
  {Nature Phys.}\ }\textbf {\bibinfo {volume} {2}},\ \bibinfo {pages} {687}
  (\bibinfo {year} {2006})}\BibitemShut {NoStop}%
\bibitem [{\citenamefont {Bushmaker}\ \emph {et~al.}(2009)\citenamefont
  {Bushmaker}, \citenamefont {Deshpande}, \citenamefont {Hsieh}, \citenamefont
  {Bockrath},\ and\ \citenamefont {Cronin}}]{Bushmaker:2009}%
  \BibitemOpen
  \bibfield  {author} {\bibinfo {author} {\bibfnamefont {A.~W.}\ \bibnamefont
  {Bushmaker}}, \bibinfo {author} {\bibfnamefont {V.~V.}\ \bibnamefont
  {Deshpande}}, \bibinfo {author} {\bibfnamefont {S.}~\bibnamefont {Hsieh}},
  \bibinfo {author} {\bibfnamefont {M.~W.}\ \bibnamefont {Bockrath}}, \ and\
  \bibinfo {author} {\bibfnamefont {S.~B.}\ \bibnamefont {Cronin}},\ }\href
  {\doibase 10.1103/PhysRevLett.103.067401} {\bibfield  {journal} {\bibinfo
  {journal} {Phys. Rev. Lett.}\ }\textbf {\bibinfo {volume} {103}},\ \bibinfo
  {pages} {067401} (\bibinfo {year} {2009})}\BibitemShut {NoStop}%
\bibitem [{\citenamefont {Xiao}\ \emph {et~al.}(2010)\citenamefont {Xiao},
  \citenamefont {Nhan}, \citenamefont {Wilson},\ and\ \citenamefont
  {Fraser}}]{Xiao:2010}%
  \BibitemOpen
  \bibfield  {author} {\bibinfo {author} {\bibfnamefont {Y.-F.}\ \bibnamefont
  {Xiao}}, \bibinfo {author} {\bibfnamefont {T.~Q.}\ \bibnamefont {Nhan}},
  \bibinfo {author} {\bibfnamefont {M.~W.~B.}\ \bibnamefont {Wilson}}, \ and\
  \bibinfo {author} {\bibfnamefont {J.~M.}\ \bibnamefont {Fraser}},\ }\href
  {\doibase 10.1103/PhysRevLett.104.017401} {\bibfield  {journal} {\bibinfo
  {journal} {Phys. Rev. Lett.}\ }\textbf {\bibinfo {volume} {104}},\ \bibinfo
  {pages} {017401} (\bibinfo {year} {2010})}\BibitemShut {NoStop}%
\bibitem [{\citenamefont {Matsuda}\ \emph {et~al.}(2010)\citenamefont
  {Matsuda}, \citenamefont {Miyauchi}, \citenamefont {Sakashita},\ and\
  \citenamefont {Kanemitsu}}]{Matsuda:2010}%
  \BibitemOpen
  \bibfield  {author} {\bibinfo {author} {\bibfnamefont {K.}~\bibnamefont
  {Matsuda}}, \bibinfo {author} {\bibfnamefont {Y.}~\bibnamefont {Miyauchi}},
  \bibinfo {author} {\bibfnamefont {T.}~\bibnamefont {Sakashita}}, \ and\
  \bibinfo {author} {\bibfnamefont {Y.}~\bibnamefont {Kanemitsu}},\ }\href
  {\doibase 10.1103/PhysRevB.81.033409} {\bibfield  {journal} {\bibinfo
  {journal} {Phys. Rev. B}\ }\textbf {\bibinfo {volume} {81}},\ \bibinfo
  {pages} {033409} (\bibinfo {year} {2010})}\BibitemShut {NoStop}%
\bibitem [{\citenamefont {L\"{u}er}\ \emph {et~al.}(2009)\citenamefont
  {L\"{u}er}, \citenamefont {Hoseinkhani}, \citenamefont {Polli}, \citenamefont
  {Crochet}, \citenamefont {Hertel},\ and\ \citenamefont
  {Lanzani}}]{Luer:2009}%
  \BibitemOpen
  \bibfield  {author} {\bibinfo {author} {\bibfnamefont {L.}~\bibnamefont
  {L\"{u}er}}, \bibinfo {author} {\bibfnamefont {S.}~\bibnamefont
  {Hoseinkhani}}, \bibinfo {author} {\bibfnamefont {D.}~\bibnamefont {Polli}},
  \bibinfo {author} {\bibfnamefont {J.}~\bibnamefont {Crochet}}, \bibinfo
  {author} {\bibfnamefont {T.}~\bibnamefont {Hertel}}, \ and\ \bibinfo {author}
  {\bibfnamefont {G.}~\bibnamefont {Lanzani}},\ }\href {\doibase
  10.1038/nphys1149} {\bibfield  {journal} {\bibinfo  {journal} {Nature Phys.}\
  }\textbf {\bibinfo {volume} {5}},\ \bibinfo {pages} {54} (\bibinfo {year}
  {2009})}\BibitemShut {NoStop}%
\bibitem [{\citenamefont {Torrens}\ \emph {et~al.}(2008)\citenamefont
  {Torrens}, \citenamefont {Zheng},\ and\ \citenamefont
  {Kikkawa}}]{Torrens:2008}%
  \BibitemOpen
  \bibfield  {author} {\bibinfo {author} {\bibfnamefont {O.~N.}\ \bibnamefont
  {Torrens}}, \bibinfo {author} {\bibfnamefont {M.}~\bibnamefont {Zheng}}, \
  and\ \bibinfo {author} {\bibfnamefont {J.~M.}\ \bibnamefont {Kikkawa}},\
  }\href {\doibase 10.1103/PhysRevLett.101.157401} {\bibfield  {journal}
  {\bibinfo  {journal} {Phys. Rev. Lett.}\ }\textbf {\bibinfo {volume} {101}},\
  \bibinfo {pages} {157401} (\bibinfo {year} {2008})}\BibitemShut {NoStop}%
\bibitem [{\citenamefont {Murakami}\ \emph {et~al.}(2009)\citenamefont
  {Murakami}, \citenamefont {Lu}, \citenamefont {Kazaoui}, \citenamefont
  {Minami}, \citenamefont {Okubo},\ and\ \citenamefont
  {Maruyama}}]{Murakami:2009}%
  \BibitemOpen
  \bibfield  {author} {\bibinfo {author} {\bibfnamefont {Y.}~\bibnamefont
  {Murakami}}, \bibinfo {author} {\bibfnamefont {B.}~\bibnamefont {Lu}},
  \bibinfo {author} {\bibfnamefont {S.}~\bibnamefont {Kazaoui}}, \bibinfo
  {author} {\bibfnamefont {N.}~\bibnamefont {Minami}}, \bibinfo {author}
  {\bibfnamefont {T.}~\bibnamefont {Okubo}}, \ and\ \bibinfo {author}
  {\bibfnamefont {S.}~\bibnamefont {Maruyama}},\ }\href {\doibase
  10.1103/PhysRevB.79.195407} {\bibfield  {journal} {\bibinfo  {journal} {Phys.
  Rev. B}\ }\textbf {\bibinfo {volume} {79}},\ \bibinfo {pages} {195407}
  (\bibinfo {year} {2009})}\BibitemShut {NoStop}%
\bibitem [{\citenamefont {Matsunaga}\ \emph {et~al.}(2010)\citenamefont
  {Matsunaga}, \citenamefont {Matsuda},\ and\ \citenamefont
  {Kanemitsu}}]{Matsunaga:2010}%
  \BibitemOpen
  \bibfield  {author} {\bibinfo {author} {\bibfnamefont {R.}~\bibnamefont
  {Matsunaga}}, \bibinfo {author} {\bibfnamefont {K.}~\bibnamefont {Matsuda}},
  \ and\ \bibinfo {author} {\bibfnamefont {Y.}~\bibnamefont {Kanemitsu}},\
  }\href {\doibase 10.1103/PhysRevB.81.033401} {\bibfield  {journal} {\bibinfo
  {journal} {Phys. Rev. B}\ }\textbf {\bibinfo {volume} {81}},\ \bibinfo
  {pages} {033401} (\bibinfo {year} {2010})}\BibitemShut {NoStop}%
\end{thebibliography}%

\end{document}